# Density, porosity, mineralogy, and internal structure of cosmic dust and alteration of its properties during high velocity atmospheric entry

T. Kohout[1, 2], A. Kallonen[1], J.-P. Suuronen[1], P. Rochette[3], A. Hutzler[3], J. Gattacceca[3, 4], D. D. Badjukov[5], R. Skála[2], V. Böhmová[2] and J. Čuda[6]

1.  Department of Physics, University of Helsinki, Finland (tomas.kohout@helsinki.fi)
2.  Institute of Geology, Academy of Sciences of the Czech Republic, Prague, Czech Republic
3.  Aix-Marseille Université/CNRS, CEREGE UM34, 13545 Aix-en-Provence, France
4.  Department of Earth, Atmospheric, and Planetary Sciences, Massachusetts Institute of Technology, 77 Massachusetts Avenue, Cambridge, MA 02139, USA
5.  V. I. Vernadsky Institute of Geochemistry and Analytical Chemistry RAS, Moscow, Russia
6.  Regional Centre of Advanced Technologies and Materials, Departments of Physical Chemistry and Experimental Physics, Palacky University Olomouc, Czech Republic

## Abstract

X-ray microtomography (XMT), X-ray diffraction (XRD) and magnetic hysteresis measurements were used to determine micrometeorite internal structure, mineralogy, crystallography, and physical properties at ~μm resolution. The study samples include unmelted, partially melted (scoriaceous) and completely melted (cosmic spherules) micrometeorites. This variety not only allows comparison of the mineralogy and porosity of these three micrometeorite types, but also reveals changes in meteoroid properties during atmospheric entry at various velocities. At low entry velocities, meteoroids do not melt, and their physical properties do not change. The porosity of unmelted micrometeorites varies considerably (0-12%) with one friable example having porosity around 50%. At higher velocities, the range of meteoroid porosity narrows, but average porosity increases (to 16-27%) due to volatile evaporation and partial melting (scoriaceous phase). Metal distribution seems to be mostly unaffected at this stage. At even higher entry velocities, complete melting follows the scoriaceous phase. Complete melting is accompanied by metal oxidation and redistribution, loss of porosity ($1 \pm 1\%$), and narrowing of the bulk ($3.2 \pm 0.5$ g/cm$^3$) and grain ($3.3 \pm 0.5$ g/cm$^3$) density range. Melted cosmic spherules with a barred olivine structure show an oriented crystallographic structure, whereas other subtypes do not.

## Introduction

Cosmic dust recovered on Earth in the form of interplanetary dust particles (IDPs) and micrometeorites is, together with larger meteorites, a valuable source of primitive extraterrestrial material. The cosmic dust input to Earth is in the 5-300 t/d range, depending on the method of observation (*Plane*, 2012). Micrometeorites and IDPs represent material released from asteroids and comets (e.g., *Nesvorny et al.*, 2006, 2010) into interplanetary space, material which carries important information about the composition and structure of asteroid surfaces and cometary dust. Thus, knowledge of the physical properties of cosmic dust is essential in interpreting the surface material and physical properties of asteroids and comets, as well as in modeling the atmospheric entry of cosmic dust (meteor phenomena).

The physical properties of larger, centimeter- to decimeter-sized extraterrestrial samples, meteorites, have undergone extensive study. *Consolmagno et al.* (2008) reviewed their bulk and grain density and porosity measured mostly using gas pycnometry and the Archimedean glass bead method.

Smaller, millimeter- to micrometer-sized interplanetary dust particles and micrometeorites have received much less study. Traditional measurement methods used on meteorites cannot be applied to these particles due to their small size. For this reason, little is known about the physical properties of cosmic dust.

Early research deduced the particle size and density of interplanetary dust from the morphology of micrometeorite impact craters on lunar rocks (e.g., *Brownlee et al.*, 1976, *Nagel et al.*, 1976). Either scanning electron microscopy (SEM) or synchrotron x-ray fluorescence (SXRF) later served to determine density and porosity of recovered IDPs and micrometeorites. Shape and volume were deduced from 2D or stereo images, and mass was usually deduced from observed mineral grain sizes and their known mineralogical density or from SXRF data. Based on these methods, the average bulk density of cosmic dust is typically around 2 g/cm$^3$ with most values between 0.6 and 5.5 g/cm$^3$. *Flynn and Sutton* (1991) reported a bimodal distribution (averaging 1-2 g/cm$^3$ and 2.7 g/cm$^3$) for interplanetary dust. However, *Love et al.* (1994) reported a unimodal distribution averaging around 2 g/cm$^3$. Few dust particles have been found to have bulk densities as low as 0.3 and as high as 6.2 g/cm$^3$. While the mid-upper range of these values is similar to the bulk densities of meteorites, values below 1 g/cm$^3$ are significantly lower, indicating porosities exceeding 50%. Published porosity values of IDPs show similarly wide variation. While *Corrigan et al.* (1997) reported porosity values between 0 and 10% (similar to that of larger chondritic meteorites), *Rietmeijer and Nuth* (2000) report densities as low as 0.1 g/cm$^3$ for a few dust particles with porosities of up to 90%.

Recently new methods such as X-ray microtomography (XMT, also known in the literature as μ-CT) have been employed (*Okazawa et al.*, 2002, *Tsuchiyama et al.*, 2004, *Taylor et al.*, 2011). XMT scans provide fully 3D volumetric X-ray attenuation maps of both the exterior and interior of the sample, providing a significant improvement over 2D SEM imaging. However, XMT resolution with the majority of X-ray tube sources is limited to ~ 1 μm depending on the Ximaging geometry, the size of the source and the resolution of the detector.

In this study, we present the physical properties, internal structure, and mineralogy derived from XMT scans and X-ray diffraction (XRD) measurements of 32 micrometeorites of various melted, partially melted and unmelted types. By comparing their properties, we evaluate changes in the physical properties of micrometeoroids as a function of their atmospheric velocity.

## Study material

Micrometeorites can be recovered from specific areas where they are concentrated or where they accumulate. Favorable concentration areas include sediments formed by melting ice caps (e.g., *Maurette et al.*, 1987) or moraine sediments (e.g., *Harvey and Maurette*, 1991). Areas of favorable accumulation include stable surfaces in desert settings, whether in Antarctica (*Rochette et al.*, 2008) or in hot deserts (*Gattacceca et al.*, 2011). In this study, we report measurements of the physical properties of cosmic dust in the form of 32 micrometeorites collected from the northern ice cap of the Novaya Zemlya archipelago in Russia (*Badjukov et al.*, 2003) and from soils collected in the central depression of the Atacama desert (*Gattacceca et al.*, 2011). Based on the classification by *Genge et al.* (2008), the collection includes various types of micrometeorite ranging from melted (glassy, porphyritic olivine and barred olivine S-type (silicate) cosmic spherules) to partly melted (scoriaceous) to well-preserved unmelted ones (mostly fine grained).

## Instruments and methods

### XMT measurements

The XMT scans were collected at the Department of Physics, University of Helsinki with custom-built Nanotom 180 NF tomography equipment (Phoenix|x-ray Systems and Services, part of GE Measurement Systems and Solutions, Germany). Its high-voltage (20-180 kV) nanofocus X-ray tube and variable imaging geometry enable the XMT equipment to scan objects with a diameter from 10 cm to 50 μm, and with a sub-micron voxel size (the edge length of one cubic volume element) for objects smaller than 2 mm. All micrometeorites from the Atacama collection and micrometeorites 1-6 from the Novaya Zemlya collection were placed into conical plastic tubes, which were closed at one end, and secured with cotton. The plastic tubes were then glued to a sample holder and placed on a rotating stage. For the smaller micrometeorites (11-19 from the Novaya Zemlya collection), the plastic tube was omitted, and the micrometeorites were glued directly onto a carbon fiber sample holder with acetone-soluble glue. We measured one micrometeorite at a time and moved the rotation stage as close to the X-ray source as possible in order to maximize the magnification. The acquisition parameters varied slightly due to differences in measurement geometry resulting from small variations in the micrometeorite mounting. The X-ray tube acceleration voltage was 80 kV, and the tube current was 180 μA. The micrometeorites were imaged over a full 360° circle with an angular step between 0.5° and 0.225° (corresponding to 720-1440 projection images). Each projection image was composed of an average of 8 to 16 transmission images, and exposure times for each transmission image varied between 500 and 2000 ms. The voxel sizes of the tomograms ranged from 0.63 μm to 0.25 μm, depending on the size of the micrometeorite. The resulting realistic resolution after 3D reconstruction is approximately 1 μm

The methodology for processing XMT data is complex. 16-bit integer (grayscale) volumes were reconstructed from the projection images with the datos|x software (Phoenix|x-ray) and digitally filtered to remove noise. On most samples, 3D median smoothing with a 3x3x3 or 5x5x5 voxel kernel was used. Gaussian smoothing with a 5x5x5 voxel kernel was used in the case of some cosmic spherules with initially good contrast and very low porosity (micrometeorites nos. 9.1 through 10.3). More sophisticated edge-preserving methods (5x5x5 voxel kernel bilateral or anisotropic diffusion filtering) were applied in the case of two unmelted micrometeorites (nos. 15 and 19) with noisier data and results were compared to the median smoothed data. It was found that the choice of filter did not significantly affect the results. The grain volume of the micrometeorites was calculated based on a manually selected gray value threshold, which separated dark (low gray value) areas representing empty space (both inside and outside the sample) from bright (high gray value) areas representing the solid micrometeorite material. One micrometeorite (no. 13) was found to contain large areas of intermediate gray value, interpreted as a partial volume effect due to pores that are smaller than the resolution of the scan. The grain volume included only a fraction of this semi-porous phase, with the proportionality constant computed linearly from the respective gray values of the three phases (bulk, semi-porous and pore space). We used two methods to calculate bulk volume. The first method incorporated a morphological closing followed by a hole-filling algorithm, which filled all pores not connected to the exterior.

The second method incorporated a convex hull algorithm. This convex hull method is suitable only for quasi-spherical objects (melted cosmic spherules or highly porous unmelted micrometeorites, such as micrometeorite 19), as it fills concave surface features and thus provides an upper limit for the bulk volume. Avizo Fire 7.0.1 software (Visualization Sciences Group) was used to compute the segmentation, morphological operations, and volume. The Pore volume was then determined as the difference between the bulk and grain volumes. Our XMT scans have 10-fold higher resolution than scans in previous XMT studies of micrometeorites, such as *Taylor et al.* (2011), and are comparable to synchrotron source-based observations by *Okazawa et al.* (2002) and *Tsuchiyama et al.* (2004, 2011).

The mass of the larger micrometeorites (cosmic spherules) was measured using Sartorius M2P microbalance at 1 μg accuracy and precision. Subsequently, bulk and grain densities were calculated from the bulk and grain volumes and mass.

*XRD measurements*

After XMT scanning, selected larger micrometeorites were investigated with XRD measurements, utilizing a second X-ray tube (IμS, Incoatec, Germany) and an area detector (Pilatus 1M, Dectris, Switzerland) incorporated into the XMT scanner. A detailed description of the combined XMT/XRD system is described in *Suuronen et al.* (2014). In brief, the XRD measurements were performed using an approximately 200 x 200 $\mu m^2$ beam of molybdenum $K_\alpha$-radiation ($\lambda$ = 0.709 Å) and an area detector in perpendicular transmission geometry. Based on the XMT reconstruction, the beam was aimed at a specified sub-volume of the micrometeorite (in most cases, the center of the micrometeorite) using the sample manipulator stage of the XMT scanner. The recorded scattering patterns were analyzed for mineralogical composition and degree of crystallite orientation in the micrometeorites with Matlab software, using the 111-diffraction peak from a silicon standard to determine the sample-to-detector distance. To sample a greater portion of the reciprocal space in the well-ordered micrometeorites, a second diffraction pattern was measured at a 90° angle to the first one.

*Electron microscopy*

To verify the compositional results derived from XMT scans, three micrometeorites were analyzed using a TESCAN VEGA 3XM scanning electron microscope (SEM) with a Bruker QUANTAX200 energy dispersive X-ray spectrometer (EDS) and CAMECA SX-100 electron probe microanalyzer (microprobe) at the Institute of Geology, Academy of Sciences of the Czech Republic.

*Magnetic properties measurements*

Hysteresis properties were measured using a Princeton Measurements Model 3900 vibrating sample magnetometer (VSM) at the Department of Physics, University of Helsinki for the Novaya Zemlya, and at CEREGE, Aix-Marseille University for the Atacama collection. Hysteresis loops were measured up to ± 1.2 T (± 1 T for the Atacama collection) field. Correction for the paramagnetic component was subsequently applied. Magnetite content was calculated from the ratio of the saturation magnetization of the micrometeorite to the saturation magnetization of the magnetite (~ 50 $Am^2/kg$).

## Results

### Micrometeorite texture and structure

XMT revealed various textures and structures of the micrometeorites studied. Melted cosmic spherules in our collection include barred and cryptocrystalline olivine or glassy subtypes (Fig. 2). Three micrometeorites (nos. 9.3, 9.10, 10.4) contained coarse-grained inclusions most likely of relict material, and one micrometeorite (no. 4) contained a large metal inclusion. One cosmic spherule with a barred structure (no. 2) was polished and analyzed with SEM-EDS. Submicron magnetite grains within the olivine lamellae were identified (Fig. 3). The size of these magnetite grains is roughly equivalent to the XMT resolution, so they show only a grainy texture in the XMT images. Weathering of the spherules due to long-term exposure on the Earth's surface, as evidenced by a selective dissolution of lamellae at the surface of the spherules (nos. 2, 5, 10.5, and 10.9) or by a rim of different density and the textured material of some of the Atacama spherules (nos. 10.3, 10.11, 10.12, 10.13, and 10.14), does not seem pervasive.

Partly melted scoriaceous micrometeorites nos. 11, 14, and 16 (Fig. 4) show a relatively homogeneous silicate matrix with spherical vesicles of various sizes. Vesicles of submicron size (below the XMT resolution) are likely to exist within the micrometeorites, so the total porosity of the micrometeorites may be even higher. Micrometeorite no. 16 contains a bright metallic phase. A polished section of this micrometeorite was prepared and observed with a microprobe (Fig. 5). The matrix consisted of enstatite and low-Ca pyroxene, and the metallic phase was identified as FeNi metal (5-10 wt% Ni). The composition of this micrometeorite is similar to that of ordinary chondrites.

In addition to rounded vesicles, micrometeorite 14 contains linear cracks and thus seems to be not only scoriaceous, but also mechanically fractured. The micrometeorite also has an apparent magnetite rim (Fig. 4).

Unmelted micrometeorites nos. 12, 13, 15, 18, and 19 (Fig. 4) show various textures and structures from compact (with metal-rich inclusions observed in micrometeorite no. 12) to highly porous and fragmented (micrometeorite no. 19). A polished section of micrometeorite 12 was analyzed using a microprobe (Fig. 6) and hydrated silicates were found to be the main micrometeorite constituents. The bright phase in the BSE (back-scattered electrons) images was identified as iron oxides. Sulfur was detected in the center of some bright phase grains, which points to their likely origin as oxidation products of troilite. In some unmelted micrometeorites we observed a magnetite rim.

*Mineralogy and crystallography*

A combination of the XRD and the magnetic hysteresis measurements permits the determination of the dominant mineralogical phases. From the XRD patterns (examples shown in Fig. 7), olivine was found to be the major constituent of all the cosmic spherules. One scoriaceous and one unmelted micrometeorite that we measured also contained olivine. The olivine compositions varied widely from Mg to Fe rich (Tables 1 and 2).

In some cosmic spherules, we observed a weak pattern of magnetite. Because we measured the XRD data in only one or two orientations, not all minor phases are necessarily detected in the XRD pattern. This is especially true for the micrometeorites with a high degree of crystallite orientation. As noted in the previous section, SEM-EDS analysis confirmed the presence of magnetite in one cosmic spherule. The measured saturation magnetization values support the presence of a ferromagnetic phase in most of the cosmic spherules, even in those in which the XRD showed no such phases (Tables 1 and 2).

One drawback of our XRD setup is its limited detection capability resulting from the measurement geometry. However, this measurement approach enables us to determine the level of preferred crystallographic orientation or randomization. An XRD pattern composed of localized point reflections is typical for a micrometeorite with a preferred orientation of its crystallographic structure. Micrometeorites with some variation in the crystalline orientation show short arcs in the XRD pattern, whereas an XRD pattern dominated by circles (resembling a powder diffraction pattern) is typical of polycrystalline micrometeorites with randomized crystallographic orientations. We observed all three types of the XRD patterns noted above (Fig. 8). In general, melted cosmic spherules with barred olivine structure show mostly preferred crystallographic orientations, whereas glassy cosmic spherules are mostly disoriented (Table 1). One partly melted scoriaceous micrometeorite showed a randomized pattern, and the one unmelted micrometeorite showed an oriented pattern (Table 2).

With cosmic spherules nos. 9.3 and 9.10 it was possible to aim X-ray beam on the inclusions within the spherule and compare the resulting XRD patterns and plots to those omitting the inclusion (Fig. 7). Fig. 9 demonstrates the measurement geometry for spherule no. 9.3. This spherule shows an overall partly oriented pattern and spherule no. 9.10 shows an overall disoriented pattern. In both cases additional weak point-like reflections (oriented pattern) related to presence of the inclusion are visible and additional peaks are observed in the identical regions of the XRD plots. It is impossible to identify corresponding mineral due to a low intensity of the inclusion signal.

*Physical properties*

The physical properties of the micrometeorites are summarized in Tables 1 and 2 together with error estimates based on repeated measurements of selected micrometeorites. The physical properties of the cosmic spherules of melted silicate (S) type are quite uniform and reveal no apparent trends with subtypes (barred, cryptocrystalline olivine or glassy) apart from the single porphyritic spherule (10.4) showing slightly higher porosity and grain density. The measured porosities were found to be low (1 ± 1%), and the bulk and grain densities were relatively uniform ($3.2 \pm 0.5$ g/cm$^3$ and $3.3 \pm 0.5$ g/cm$^3$, respectively) Only one spherule (no. 4) was found to be significantly denser (bulk and grain density 5.6 g/cm$^3$ and 5.7 g/cm$^3$, respectively) due to the presence of a large metallic inclusion.

With the partially melted, scoriaceous, and unmelted micrometeorites, it was only possible to determine the porosity. We could not determine density due to the small particle size. Scoriaceous micrometeorites have higher porosities, ranging from 16% up to 27% (Table 2). Pristine unmelted micrometeorites show wider variation in their porosity than do the previous two classes. The fine-grained unmelted micrometeorites typically have porosities in the range of 0 to 12% (Table 2). One highly porous fragmental micrometeorite (no. 19) has a porosity exceeding 50% (Fig.10). This value represents the highest porosity in our data set.

In table 2 a bulk and grain density estimate is given to one scoriaceous (no. 11) and one unmelted (no. 13) micrometeorite. The grain density is in this case inferred from observed mineralogy (4.0 g/cm$^3$ for intermediate olivine and 3.5 g/cm$^3$ for magnesium-rich olivine) and bulk density is subsequently calculated from grain density and observed porosity.

A complete set of all XMT and XRD figures and videos is available as supplementary material at: http://www.mv.helsinki.fi/kohout/Supplementary%20material/

## Discussion

The physical properties of micrometeorites fall within the range of previously published values (*Flynn and Sutton*, 1991 (unmelted IDPs), *Love et al.*, 1994 (unmelted IDPs), *Corrigan et al.*, 1997 (unmelted IDPs), *Okazawa et al.*, 2002 (scoriaceous and unmelted micrometeorites), *Tsuchiyama et al.*, 2004 (scoriaceous and unmelted micrometeorites), *Taylor et al.*, 2011 (all micrometeorite types), Tsuchiyama et al. 2011 (Itokawa regolith particles). Our micrometeorite set contains examples of melted, partially melted, and unmelted cosmic dust particles and the observed trends in porosity (low porosity of melted micrometeorites, high porosity of scoriaceous micrometeorites and high porosity variations among unmelted ones) are consistent with these studies. In this work we measured all micrometeorites with the same instrumentation and methodology, thereby enabling direct comparison of various types.

Comparing the physical properties, especially the porosity, of cosmic spherules (melted micrometeorites) to those of partially melted or unmelted meteorites enables us to evaluate changes in meteoroid properties as a function of the atmospheric entry velocity and angle (Fig. 11). Unmelted micrometeorites represent nearly pristine cosmic dust which entered the atmosphere slowly, typically between 11.2 and 15 km/s and at a rather shallow entry angle *(Love and Brownlee*, 1991). Thermal changes, if any, are limited to the presence of the magnetite rim at the perimeter of some micrometeorites (Fig. 4). Such a magnetite rim is analogous to the fusion crusts covering a pristine meteorite interior of larger meteorites. The physical properties of these micrometeorites remain nearly unchanged and are representative of the cosmic dust in the Earth's vicinity. These micrometeorites seem to be rather heterogeneous with a widely variable porosity (0-12%). This is within range of various chondritic meteorites (*Consolmagno et al.*, 2008) and thus may resemble meteorite fine-grained matrix. Highly porous micrometeorite (no. 19) with porosity around 50% represents a class of its own and its appearance resembles dust aggregate particles similar to those experimentally prepared by *Wurm and Blum*, 1998. Unfortunately we were not able to determine its mineralogy due to its small size.

Partially melted, scoriaceous micrometeorites represent dust particles that enter the atmosphere at slightly higher velocities, typically in the range of 12-20 km/s and at steeper entry angles *(Love and Brownlee*, 1991). Whereas fragmental micrometeorites such as no. 19 would likely disintegrate completely during entry under such conditions, more compact micrometeorites survive the greater heating and stress. The heat generated during meteoroid deceleration partially melts and evaporates volatiles, resulting in the growth of large vesicles within the silicate matrix. However, the extent of melting is insufficient to cause homogenization of the meteoroid, oxidation of metal (FeNi metal is preserved in the micrometeorite no. 16, Fig. 5), or to change its shape from irregular to quasi-spherical, but partial melting does result in higher micrometeorite porosity (25-30%).

In contrast, cosmic spherules (melted micrometeorites) represent dust particles entering the atmosphere at high velocities, typically over 15 km/s and at steeper angles (*Love and Brownlee*, 1991). Heating during atmospheric entry completely melts the meteoroid, causes loss of up to half of its initial mass, and changes its shape to a droplet-like quasi-sphere (*Love and Brownlee*, 1991). Meteoroid homogenization and metal oxidation into magnetite lowers and narrows the range of the bulk ($3.2 \pm 0.5$ g/cm$^3$) and grain ($3.3 \pm 0.5$ g/cm$^3$) densities and reduces the porosity. Thus, density and porosity values are lower compared to chondritic meteorites (*Consolmagno et al.*, 2008) or chondritic (LL) Itokawa regolith particles (Tsuchiyama et al. 2011). The compact nature of the cosmic spherules resembles chondrules (being melt product, however in different red-ox conditions) rather than chondritic matrix. The spherule no. 4 has higher densities (bulk and grain density 5.6 g/cm$^3$ and 5.7 g/cm$^3$, respectively) similar to stony-iron meteorites (*Consolmagno et al.*, 2008) due to presence of a large metallic inclusion.

We must therefore bear in mind that the melted cosmic spherules could originally have been more porous, larger and of a wider density range (as unmelted cosmic dust particles are). Upon atmospheric entry, their porosity most likely increases initially, when partial melting occurs (scoriaceous phase), and is subsequently lost as the micrometeorites reach an entirely melted stage and solidify into spherules. Based on modeling by *Love and Brownlee* (1991), this transformation lasts only for the first few seconds.

Our micrometeorite density and porosity values also reveal similarities to those of shower and sporadic meteoroids derived from camera observations. *Babadzhanov & Kokhirova* (2009) report grain densities ranging from 2.2 to 3.4 g/cm$^3$ and porosities ranging from 0 to 83% for shower meteoroids. *Borovička et al.* 2010 report porosities in similar range. *Babadzhanov* (2001) and *Bellot-Rubio et al.* (2002) report lower bulk density values of 0.4-2.9 g/cm$^3$ and 0.45-1.9 g/cm$^3$ respectively, requiring most likely significant porosity. Our values also most closely resemble the bulk density and porosity values of group I (3.7 g/cm$^3$, ~ 0%) and II (2.07 g/cm$^3$, ~ 50%) meteoroids described in *Ceplecha et al.* (1993) and *ReVelle* (2001). Through comparison of these values to our results we conclude that these meteors in their luminous phase represent either scoriaceous phase (meteors with densities below 3 g/cm$^3$, and significant porosity, e. g. group II) or the terminal molten phase (meteors with densities over 3.7 g/cm$^3$ and absent porosity, e. g. group This comparison is valid for meteoroids of similar size to the studied micrometeorites.

**Conclusions**

The described XMT, XRD, and magnetic hysteresis methodology enables imaging of the internal structure of micrometeorites, as well as determination of the mineralogy, crystallography, and physical properties at ~µm resolution. Our study sample set consists of unmelted, partially (scoriaceous) and completely melted (cosmic spherules) micrometeorites, which allows comparison of the mineralogy and porosity of these three types of micrometeorites, as well as revealing changes in meteoroid properties during atmospheric entry at various velocities. At low velocities, the meteoroid does not melt, and its physical properties do not change significantly. At higher velocities, meteoroid porosity increases due to the evaporation of volatiles and partial melting (scoriaceous phase). Metal distribution seems largely unaffected at this stage. At even higher velocities, complete melting follows the scoriaceous phase, accompanied by metal oxidation and redistribution, loss of porosity, and narrowing of the density range. Melted cosmic spherules with a barred olivine structure show a preferred orientation in their crystallographic structure, whereas glassy cosmic spherules are mostly disoriented. Inclusions present in some cosmic spherules show an oriented crystallographic structure. The porosity behavior (initial increase followed by total loss during high velocity entry) is an especially important fact for consideration in modeling meteor phenomena.

**Acknowledgements**

We would like to thank to the reviewers Daniel Britt and Joe Friedrich for constructive comments on the manuscript and to associate editor Mike Zolensky for editorial handling. This work is supported by Academy of Finland project No. 257487 and Ministry of Education, Youth and Sports of the Czech Republic grant No. LH12079. We also thank for the support by the Operational Program Research and Development for Innovations – European Regional Development Fund (CZ.1.05/2.1.00/03.0058) and Operational Program Education for Competitiveness – European Social Fund (CZ.1.07/2.3.00/20.0017, CZ.1.07/2.3.00/20.0170, CZ.1.07/2.3.00/20.0155, and CZ.1.07/2.3.00/20.0056) of the Ministry of Education, Youth and Sports of the Czech Republic.

Table 1: List of cosmic spherules with their physical properties, mineralogy and tentative classification. Approximate size is a rough approximation of the mean diameter. Mass is given with 1µg accuracy and bulk volume is given up to two significant digits. CS – cosmic spherule, BO – barred olivine, G – glass, PO – porphyritic olivine, rel – containing relict grain(s), NZ – Novaya Zemlya, AT – Atacama, $V_B$ – bulk volume, $\rho_B$ – bulk density, $\rho_G$ – grain density, p – porosity, Ol (Fe) – iron-rich olivine, Ol (Mg) – magnesium-rich olivine, Ol (int.) – intermediate olivine, Mt – magnetite, O – preferred orientation of the crystallographic structure, PO – partly oriented crystallographic structure, RO – randomized crystallographic structure, n.d. – not determined.

| Micrometeorite number | Type | Origin | Approximate size (µm) | Mass (µg) | $V_B$ ($10^6 \mu m^3$) | $\rho_B$ (g/cm$^3$) | $\rho_G$ (g/cm$^3$) | p (%) | Mineralogy major | Mineralogy minor | $M_s$ (Am$^2$/kg) | Mt (wt%) | Structure |
|---|---|---|---|---|---|---|---|---|---|---|---|---|---|
| 1 | S-BO | NZ | 320 | 53 | 16 | 3.5 | 3.6 | 1 | Ol (Mg) | Mt | | | O |
| 2 | S-BO | NZ | 390 | 86 | 24 | 3.6 | 3.7 | 2 | - | - | 9.0 | 18 | - |
| 3 | S-BO | NZ | 170 | 93 | 28 | 3.4 | 3.4 | 1 | - | - | 8.2 | 16 | - |
| 4 | S-PO | NZ | 190 | 181 | 32 | 5.6 | 5.7 | 0 | - | - | 22.7 | 45 | - |
| 5 | S-G | NZ | 330 | 61 | 19 | 3.3 | 3.4 | 3 | - | - | 5.1 | 10 | - |
| 6 | S-G | NZ | 450 | 125 | 38 | 3.3 | 3.4 | 1 | - | - | 5.6 | 11 | - |
| 9.1 | S-PO | AT | 630 | 326 | 110 | 3.0 | 3.0 | 0 | Ol (int.) | n. d. | 4.0 | 8 | RO |
| 9.2 | S-BO | AT | 530 | 235 | 73 | 3.2 | 3.2 | 1 | Ol (Mg) | Mt | 8.1 | 16 | O |
| 9.3 | S-BO, rel | AT | 850 | 489 | 170 | 2.9 | 2.9 | 1 | Ol (int.) | Mt | 5.0 | 10 | PO |
| 9.9 | S-BO | AT | 550 | 254 | 81 | 3.1 | 3.1 | 0 | Ol (Mg) | Mt | 5.8 | 12 | O |
| 9.10 | S-G, rel | AT | 590 | 258 | 87 | 3.0 | 3.0 | 1 | Ol (int.) | n. d. | 2.0 | 4 | RO |
| 10.2 | S-BO | AT | 610 | 254 | 83 | 3.0 | 3.0 | 0 | Ol (int.) | n. d. | 2.8 | 6 | O |
| 10.3 | S-BO | AT | 670 | 353 | 120 | 3.1 | 3.1 | 0 | Ol (Fe) | Mt | 8.3 | 17 | O |
| 10.4 | S-PO, rel | AT | 720 | 501 | 166 | 3.0 | 3.2 | 5 | Ol (int.) | Mt | 4.8 | 10 | O |
| 10.5 | S-BO | AT | 670 | 398 | 124 | 3.2 | 3.2 | 1 | Ol (int.) | Mt | 8.4 | 17 | O |
| 10.6 | S-BO | AT | 710 | 310 | 102 | 3.0 | 3.0 | 0 | Ol (int.) | Mt | 9.6 | 19 | O |
| 10.8 | S-BO | AT | 630 | 265 | 80 | 3.3 | 3.3 | 0 | Ol (Fe) | Mt | 15.8 | 32 | O |
| 10.9 | S-BO | AT | 680 | 338 | 110 | 3.1 | 3.1 | 0 | Ol (int.) | Mt | 10.1 | 20 | O |
| 10.10 | S-G | AT | 550 | 247 | 82 | 3.0 | 3.0 | 0 | Ol (int.) | Mt | 3.7 | 7 | RO |
| 10.11 | S-G | AT | 670 | 292 | 102 | 2.9 | 2.9 | 0 | Ol (int.) | Mt | 7.4 | 15 | RO |
| 10.12 | S-G | AT | 600 | 300 | 97 | 3.1 | 3.1 | 1 | Ol (int.) | Mt | 5.0 | 10 | RO |
| 10.13 | S-G | AT | 700 | 405 | 135 | 3.0 | 3.0 | 1 | Ol (int.) | Mt | 2.0 | 4 | RO |
| 10.14 | S-G | AT | 630 | 266 | 91 | 2.9 | 2.9 | 0 | Ol (int.) | Mt | 5.1 | 10 | PO |
| 10.15 | S-BO | AT | 600 | 236 | 81 | 2.9 | 2.9 | 0 | Ol (int.) | Mt | 11.0 | 22 | O |
| Error | | | | ± 1 | ± 1 | ± 0.1 | ± 0.1 | ± 3 | | | ± 0.1 | ± 1 | |

Table 2: List of partially melted and unmelted micrometeorites with their physical properties, mineralogy and tentative classification. Approximate size is a rough approximation of the mean diameter. Bulk volume is given up to two significant digits. PM – partially melted, UM – unmelted, Sc – scoriaceous, Fg – fine grained, P – porous, NZ – Novaya Zemlya, $V_B$ – bulk volume, $\rho_B$ – bulk density, $\rho_G$ – grain density, p – porosity, Ol (Mg) – magnesium-rich olivine, Ol (int.) – intermediate olivine, Mt – magnetite, O – preferred orientation of the crystallographic structure, RO – randomized crystallographic structure, n.d. – not detected.

| Micrometeorite number | Type | Origin | Approximate size (μm) | Mass (μg) | $V_B$ ($10^6 \mu m^3$) | $\rho_B$ (g/cm$^3$) | $\rho_G$ (g/cm$^3$) | p (%) | Mineralogy major | Mineralogy minor | Structure |
|---|---|---|---|---|---|---|---|---|---|---|---|
| 11 | PM-Sc | NZ | 230 | - | 1.6 | 3.0 | 4.0 | 25 | Ol (int.) | n. d. | RO |
| 14 | PM-Sc | NZ | 160 | - | 1.3 | - | - | 27 | - | - | - |
| 16 | PM-Sc | NZ | 170 | - | 1.4 | - | - | 16 | - | - | - |
| 12 | UM-Fg | NZ | 210 | - | 2.8 | - | - | 0 | - | - | - |
| 13 | UM-Fg | NZ | 230 | - | 1.0 | 3.1 | 3.5 | 12 | Ol (Mg) | Mt | O |
| 15 | UM-Fg | NZ | 170 | - | - | - | - | 0 | - | - | - |
| 18 | UM-Fg | NZ | 120 | - | 0.55 | - | - | 9 | - | - | - |
| 19 | UM-P | NZ | 150 | - | 1.2 | - | - | 51 | - | - | - |
| Error | | | | | ± 0.2 | ± 0.3 | ± 0.3 | ± 5 | | | |

Fig. 1: The X-ray microtomography setup.

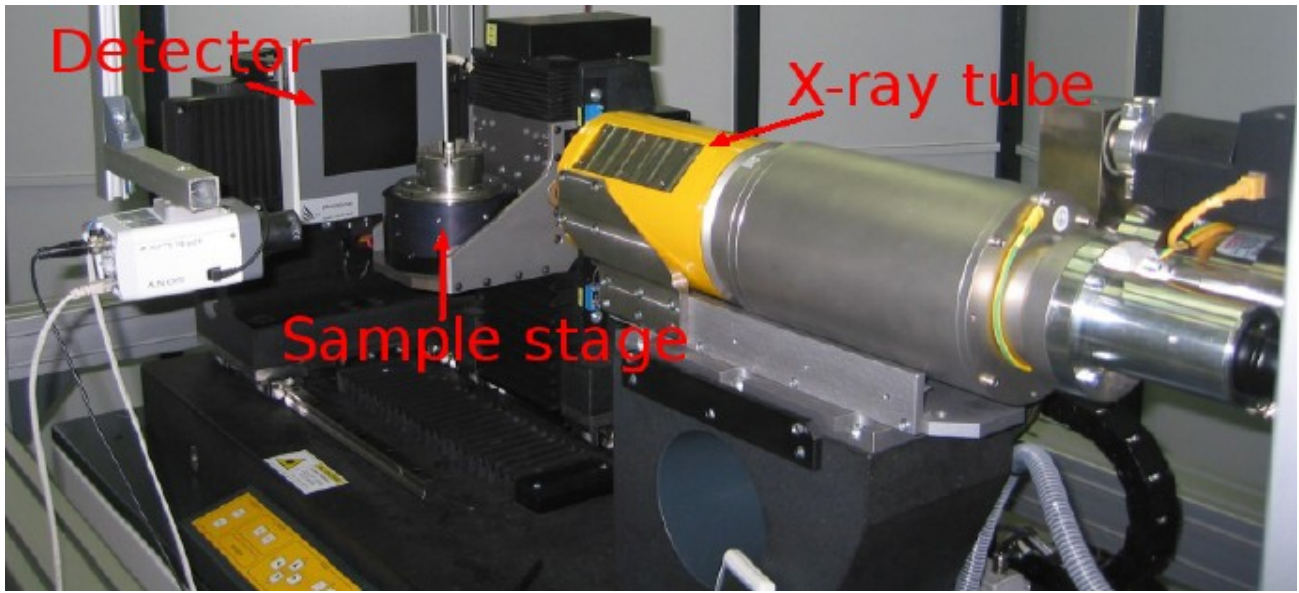

Fig. 2: Tomography sections of entirely melted micrometeorites (cosmic spherules). The weathering effects can be seen on some micrometeorites (partial dissolution extending from the surface on spherules nos. 2, 5, 10.5 and 10.9; weathering rim on spherules no. 10.3, 10.11, 10.12, 10.13 and 10.14). Micrometeorite size and origin is indicated in Table 1. Spherules nos. 9.3, 9.10, and 10.4 contain an inclusion – probably composed of pristine material.

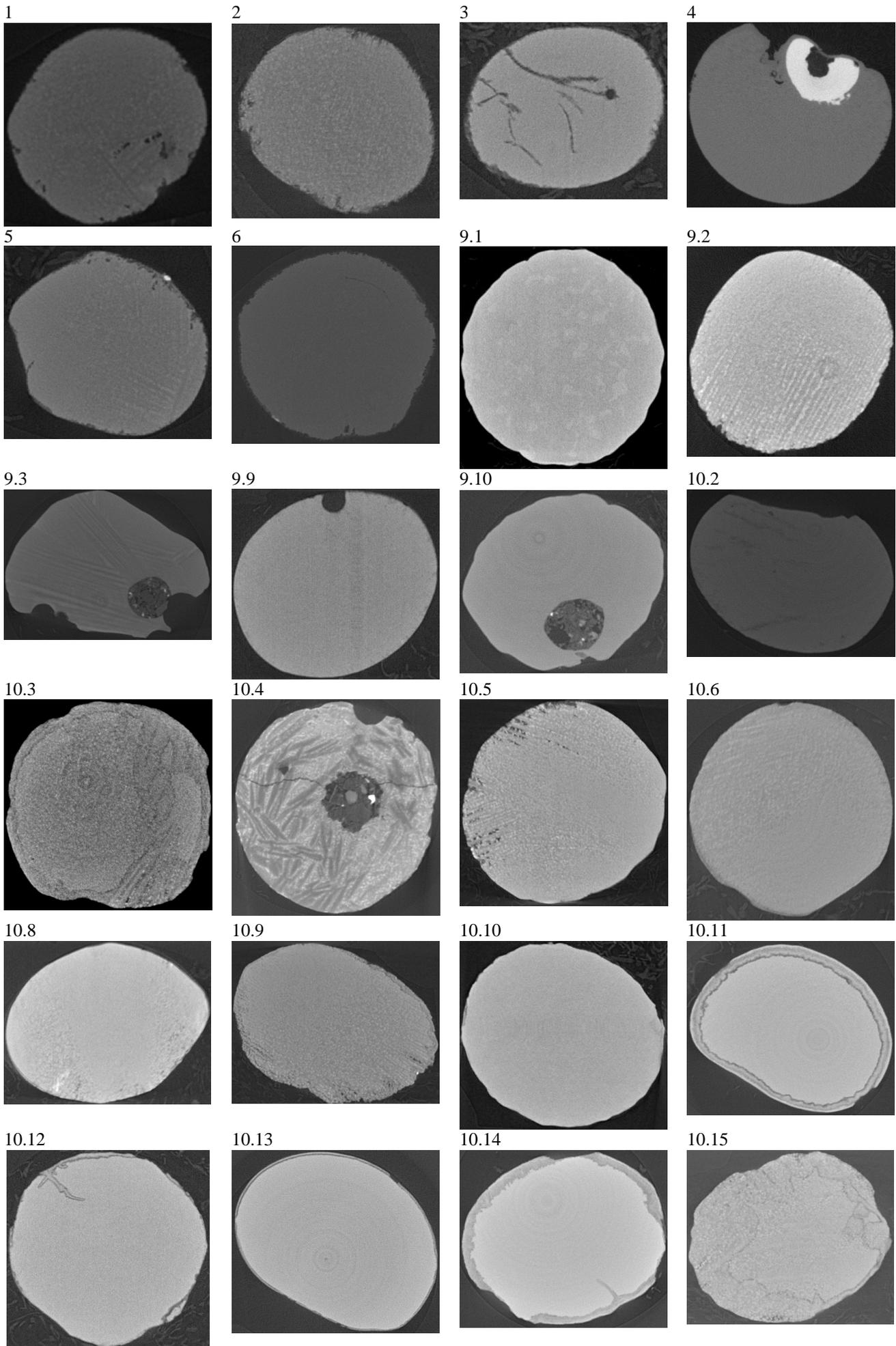

Fig. 3: Comparison of the cosmic spherule no. 2 tomography section (left) to electron microscope (backscattered electrons) image (right) of sample 2 (sections are not of the same orientation). The barred olivine structure with submicron magnetite grains can be observed.

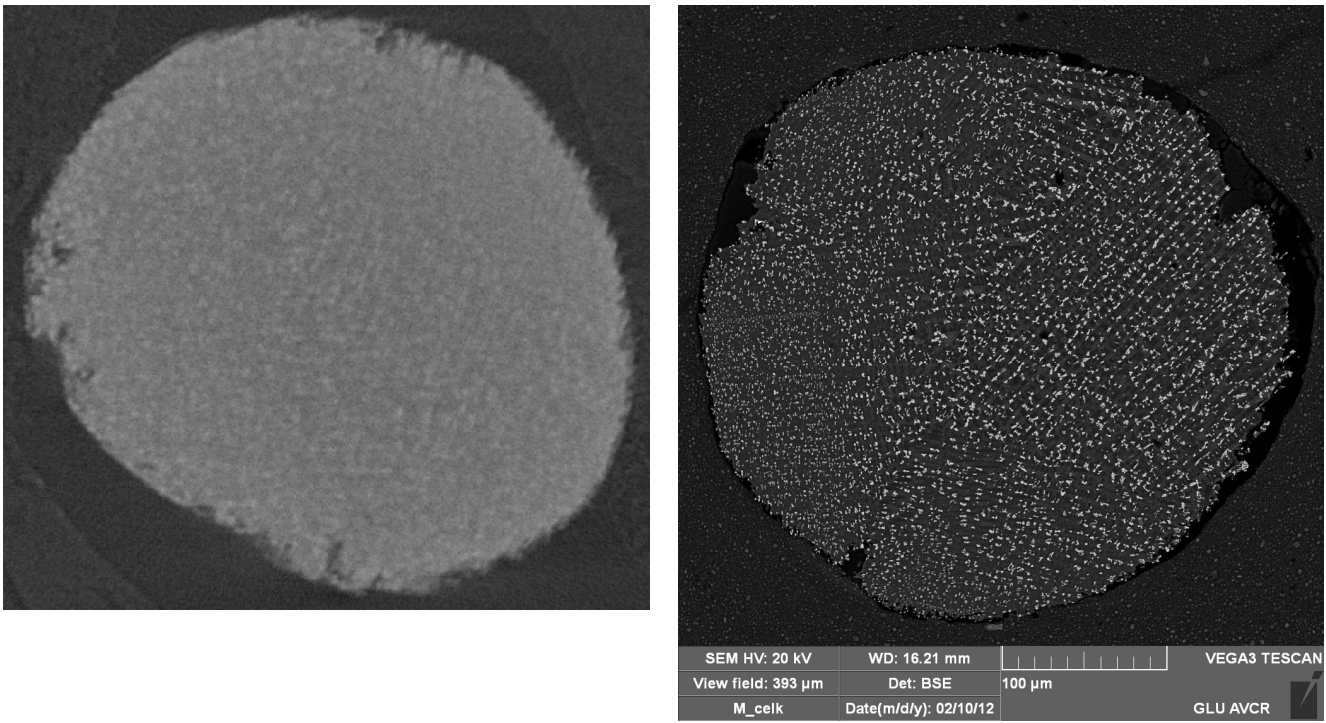

Fig.4: Tomography sections of the partially melted (scoriaceous) micrometeorites nos. 11, 14, and 16 and unmelted micrometeorites nos. 12, 13, 15, 18, and 19. Micrometeorite size and origin is indicated in Table 2.

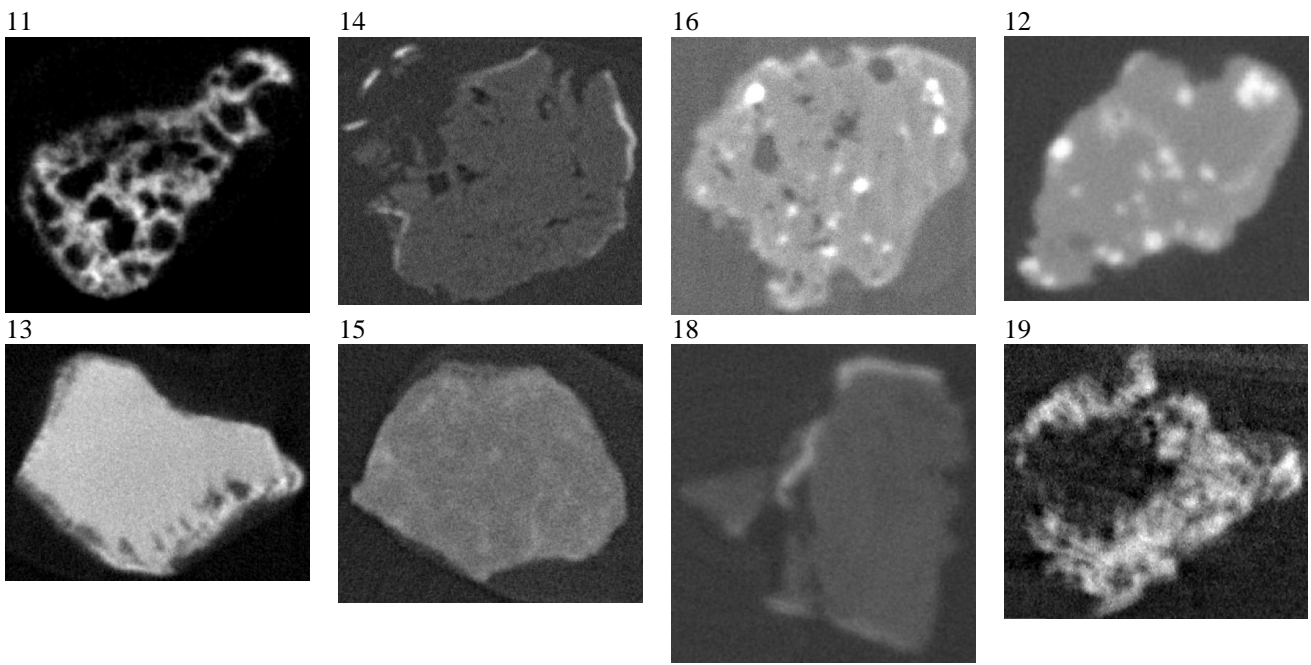

Fig. 5: Comparison of the micrometeorite no. 16 tomography section (left) to electron microscope (backscattered electrons) image (right. sections are not of the same orientation). The matrix is composed of enstatite and low-Ca pyroxene while the bright fraction is FeNi metal. The magnetite rim surrounding the sample perimeter can be also seen.

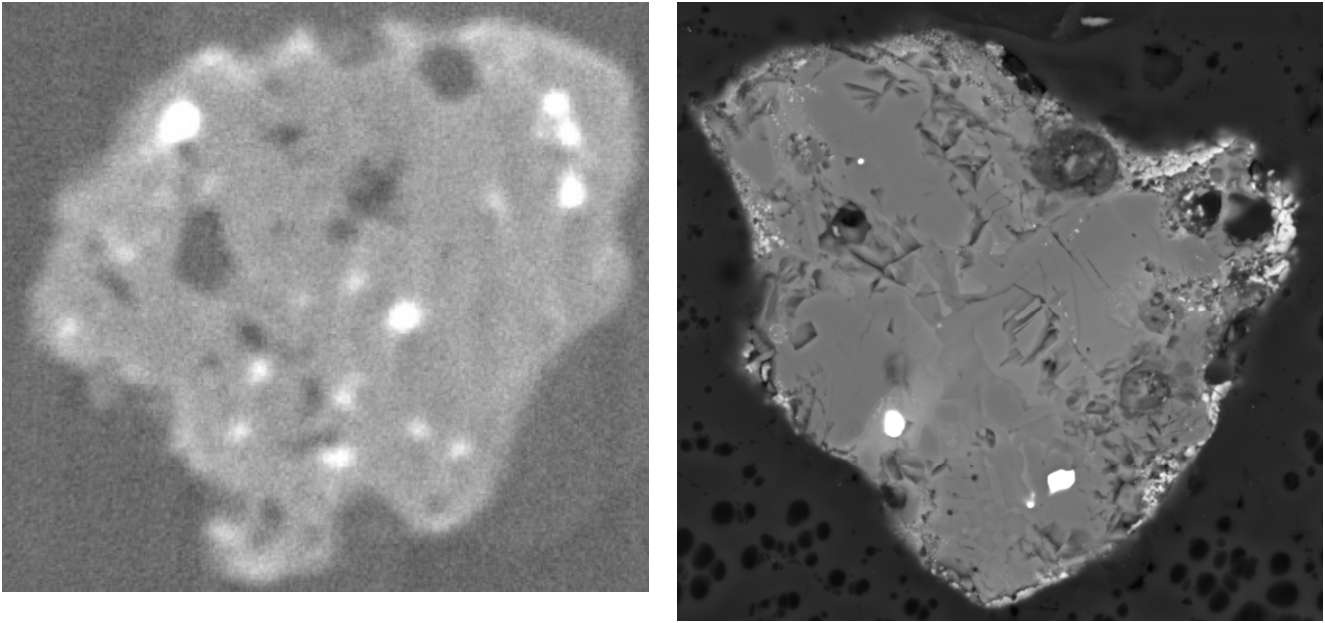

Fig. 6: Comparison of the micrometeorite no. 12 tomography section (left) to electron microscope (backscattered electrons) image (right, sections are not of the same orientation). The matrix is composed of hydrated silicates while the bright fraction is iron oxides. Inset: In some of the oxide grains sulfur can be found in their center (Fe: green; S: yellow, Si: blue).

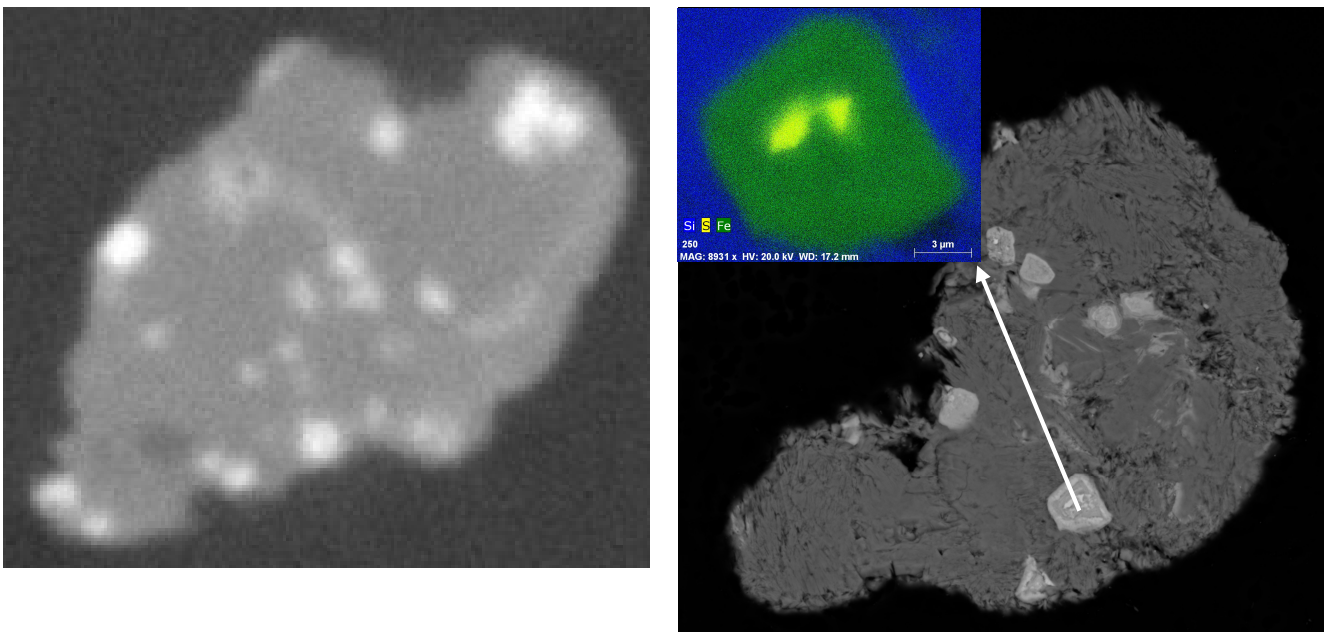

Fig. 7: Comparison of the cosmic spherules nos. 9.3 and 9.10 X-ray diffraction patterns and plots focused on the inclusion (left) to the ones offset from it (right). Additional point-like reflections originating from the inclusion are marked with arrows.

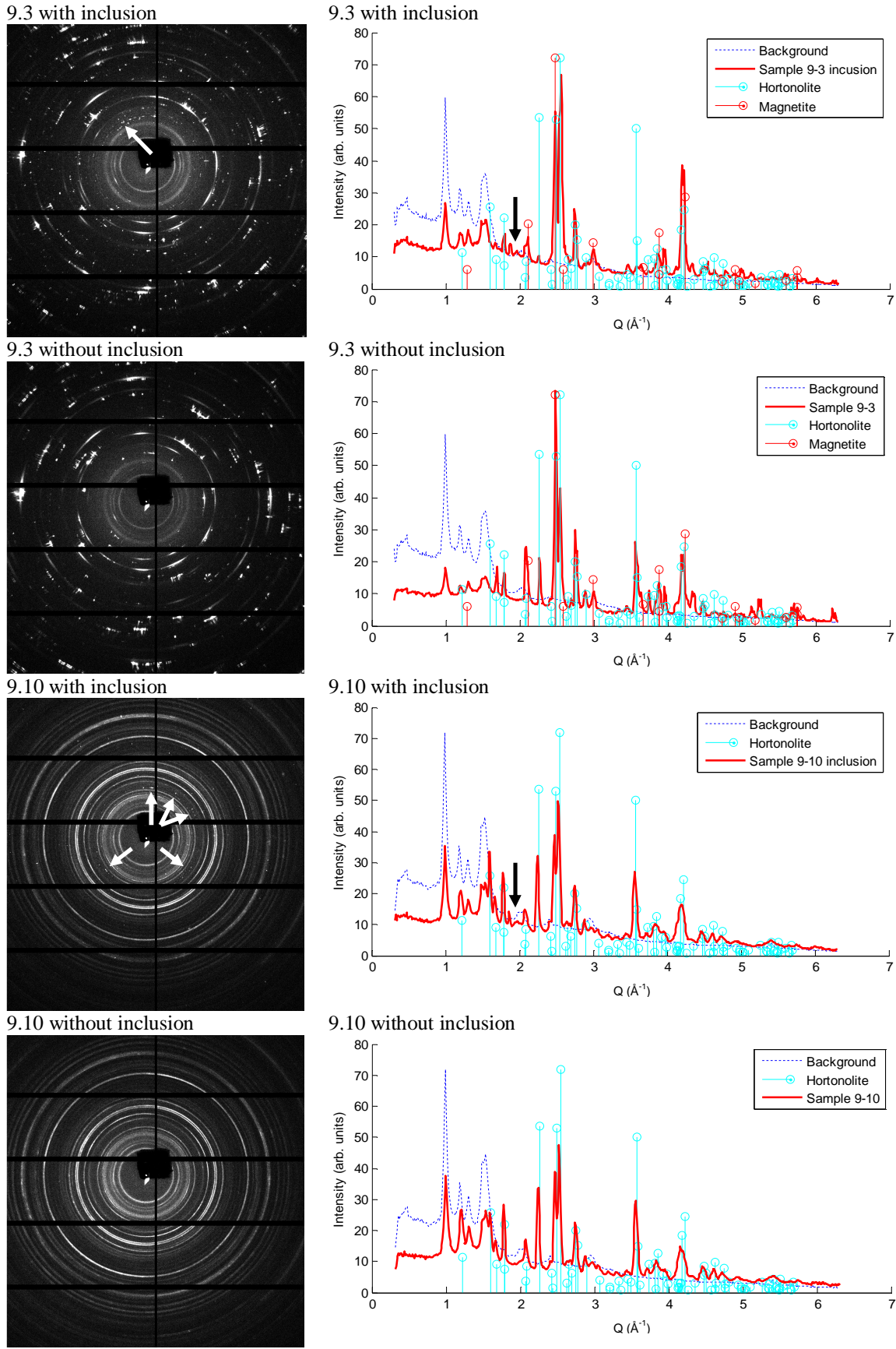

Fig. 8: Examples of X-ray diffraction patterns of cosmic spherules with preferred orientation of the crystallographic structure (no. 9.2), partly oriented crystallographic structure (no. 9.3), and randomized crystallographic orientations (no. 9.1).

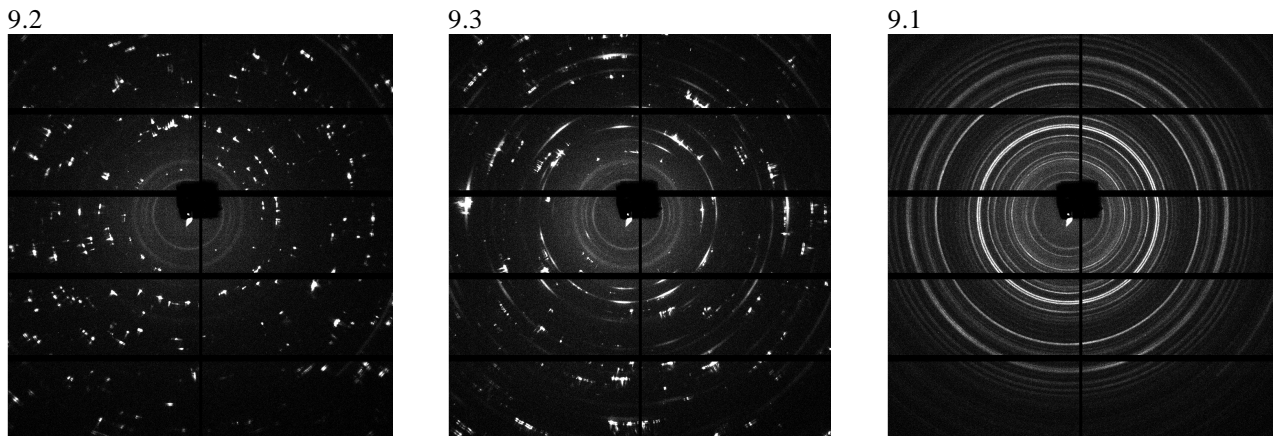

Fig. 9: The measurement geometry of the cosmic spherule no. 9.3 with X-ray paths aimed at and omitting the inclusion.

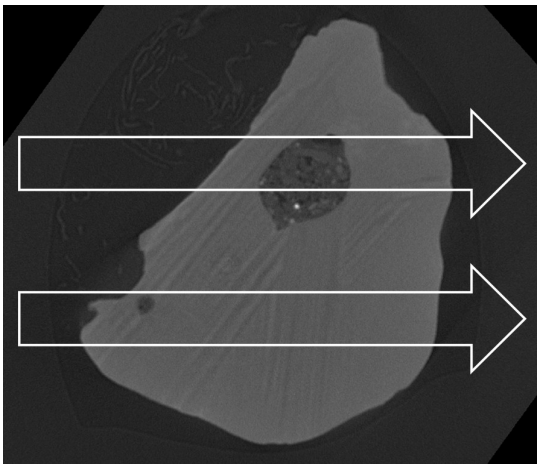

Fig. 10: Surface rendering of the fragmental unmelted micrometeorite no. 19.

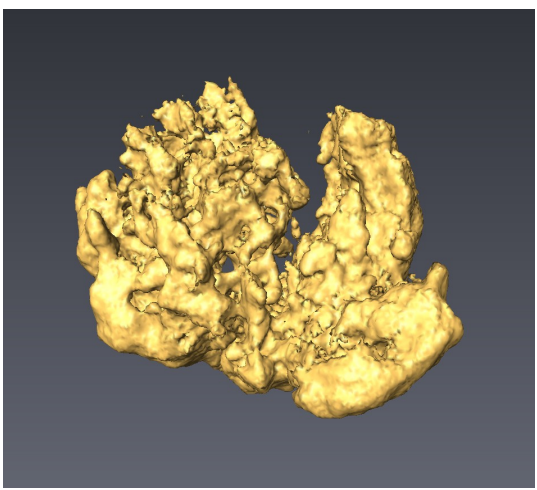

Fig. 11: Evolution of meteoroid physical properties as a function of its entry velocity.

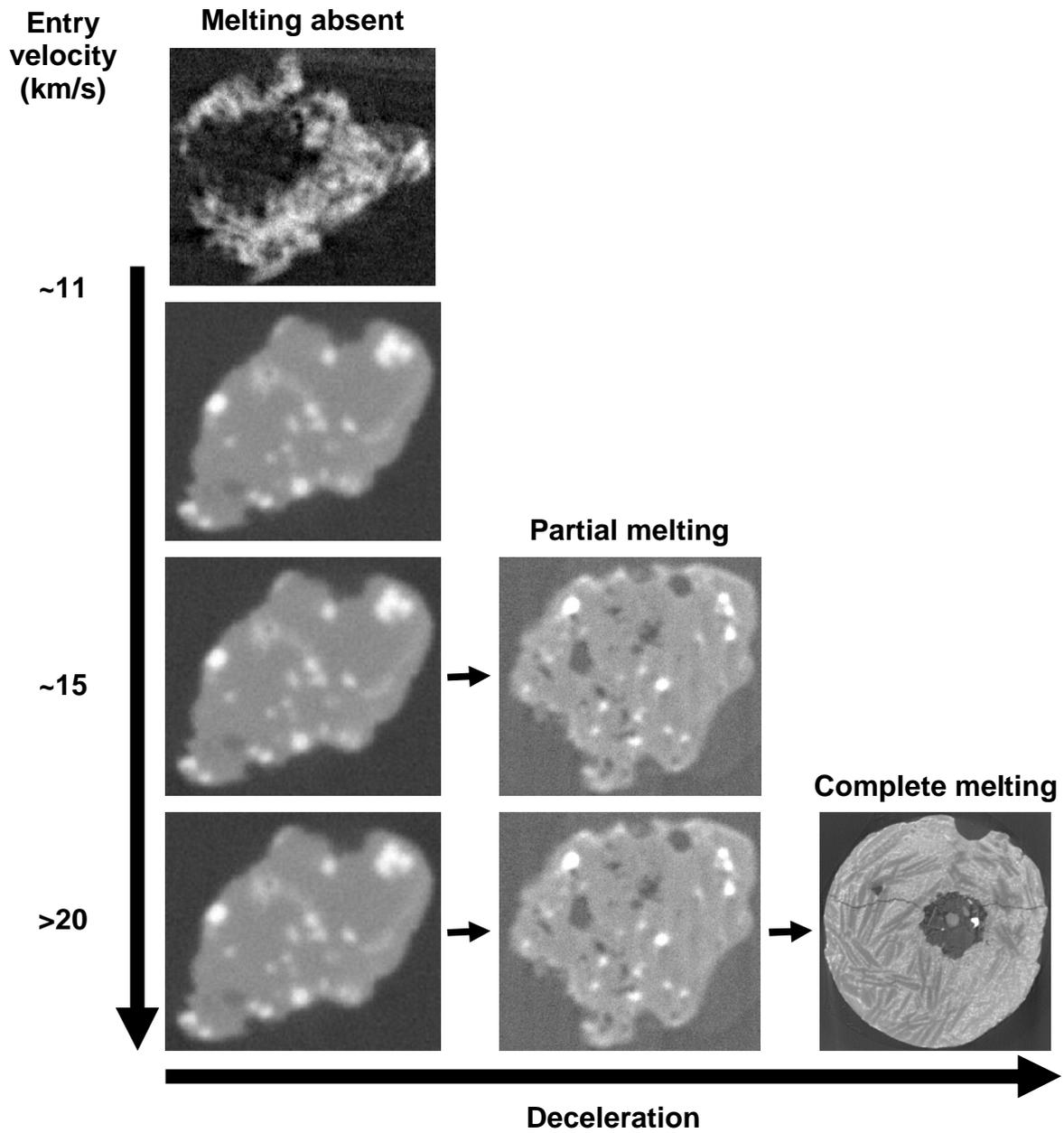